\def\aa{{A\&A}}

\def\aj{{AJ}}
\def\annrev{{ARA\&A}}
\def\apj{{ApJ}}
\def\apjs{{ApJS}}

\def\mnras{{MNRAS}}
\def\nat{{Nature}}
\def\pasp{{PASP}}
\def\eg{e.g.,\ }
\def\ie{i.e.,\ }
\def\etal{et~al.\ }
\def\kms{~km~s$^{-1}$}

\documentclass[cup5b]{caps}
\usepackage{graphicx}
\usepackage{amssymb}
\usepackage{ociwsymp3}   

\HeadText{J. C. Mihos}

\begin{document}

\pagenumbering{arabic}

\author[]{J. CHRISTOPHER MIHOS\\Case Western Reserve University}

\chapter{Interactions and Mergers of \\ Cluster Galaxies}

\begin{abstract}

The high-density environment of galaxy clusters is ripe for
collisional encounters of galaxies. While the large velocity
dispersion of clusters was originally thought to preclude slow
encounters, the infall of smaller groups into the cluster environment
provides a mechanism for promoting slow encounters and even mergers
within clusters. The dynamical and star-forming response of galaxies
to a close encounter depends on both their internal structure and on
the collisional encounter speed --- fast encounters tend to trigger modest, 
disk-wide responses in luminous spirals, while slow encounters are more able to
drive instabilities that result in strong nuclear activity. While the
combined effects of the cluster tidal field and ram pressure stripping
make it difficult for individual cluster galaxies to participate in
many merger-driven evolutionary scenarios, infalling groups represent
a natural site for these evolutionary processes and may represent a
``preprocessing'' stage in the evolution of cluster
galaxies. Meanwhile, the efficiency of tidal stripping also drives the
formation of the diffuse intracluster light in galaxy clusters;
deep imaging of clusters is beginning to reveal evidence for
significant substructure in the intracluster light.

\end{abstract}

\section{Interactions of Cluster Galaxies}

The importance of collisions in the life of cluster galaxies can be
seen through a simple rate argument. A characteristic number of
interactions per galaxy can be written as $N \approx n\sigma \upsilon 
t$, where $n$ is the number density of galaxies in a cluster, $\sigma$
is the cross section for interactions, $\upsilon$ is the encounter velocity,
and $t$ is the age of the cluster. If $\sigma = \pi r_p^2$, where $r_p$ is
the impact parameter, and $\upsilon=\sqrt 2 \sigma_\upsilon$, then for a cluster like Coma
we have 
$$
N \ \approx\ 4\ \left({n\over {250\ {\rm Mpc}^{-3}}}\right)
            \left({r_p\over {20\ {\rm kpc}}}\right)^2
	    \left({\sigma_\upsilon \over {1000\ {\rm ~km~s^{-1}}}}\right)
	    \left({t\over{\rm 10\ Gyr}}\right).
$$
While very crude, this calculation shows that it is reasonable to expect
that over the course of its lifetime in the cluster, a typical galaxy
should experience several close interactions with other cluster members.

While interactions should be common in clusters, they will also be
fast. Because the characteristic encounter velocity is much higher
than the typical circular velocities of galaxies, these perturbations
will be impulsive in nature. Simple analytic arguments suggest that
both the energy input and dynamical friction should scale as $\upsilon^{-2}$
(Binney \& Tremaine 1987), so that a fast encounter does less damage
and is much less likely to lead to a merger than are the slow
encounters experienced by galaxies in the field. A common view has
arisen, therefore, that slow interactions and mergers of galaxies are
a rarity in massive clusters (\eg Ostriker 1980), and that much of the
dynamical evolution in cluster galaxy populations is driven by the
effects of the global tidal field (\eg Byrd \& Valtonen 1990;
Henriksen \& Byrd 1996).  Combined with the possible effects of ram
pressure stripping of the dense interstellar mediuum (ISM) (Gunn \& Gott 1972) 
or hot gas in galaxy halos (``strangulation''; Larson, Tinsley, \& Caldwell 
1980), a myriad of processes, aside from galaxy interactions themselves, 
seemed available to transform cluster galaxies.

However, the abandonment of individual collisions as a mechanism to
drive cluster galaxy evolution has proved premature.  More recent work
on the dynamical evolution of cluster galaxies has emphasized the importance 
of fast collisions. Moore \etal (1996) and Moore, Lake, \& Katz (1998) have 
shown that
{\it repeated} fast encounters, coupled with the effects of the global
tidal field, can drive a very strong response in cluster galaxies. For
galaxy-like potentials, the amount of heating during an impulsive
encounter scales like $\Delta E/E \sim r_p^{-2}$, such that distant
encounters impart less energy.  However, this effect is balanced by
the fact that under simple geometric weighting the number of
encounters scales as $r_p^2$, so that the total heating, summed over
all interactions, can be significant. While this simple argument
breaks down when one considers more realistic galaxy potentials and
the finite time scales involved, in hindsight it is not surprising that
repeated high-speed encounters --- ``galaxy harassment'' ---
should drive strong evolution.  However, the efficacy of harassment is
largely limited to low-luminosity hosts, due to their slowly rising
rotation curves and low-density cores.  In luminous spirals, the
effects of harassment are much more limited (Moore \etal 1999). This
effect leads to a situation where harassment can effectively describe
processes such as the formation of dwarf ellipticals (Moore \etal
1998), the fueling of low-luminosity AGNs (Lake, Katz, \& Moore 1998), and the
destruction of low-surface brightness galaxies in clusters (Moore
\etal 1999), but is less able to explain the evolution of luminous
cluster galaxies.

\begin{figure*}[t]
\includegraphics[width=1.00\columnwidth,angle=0,clip]{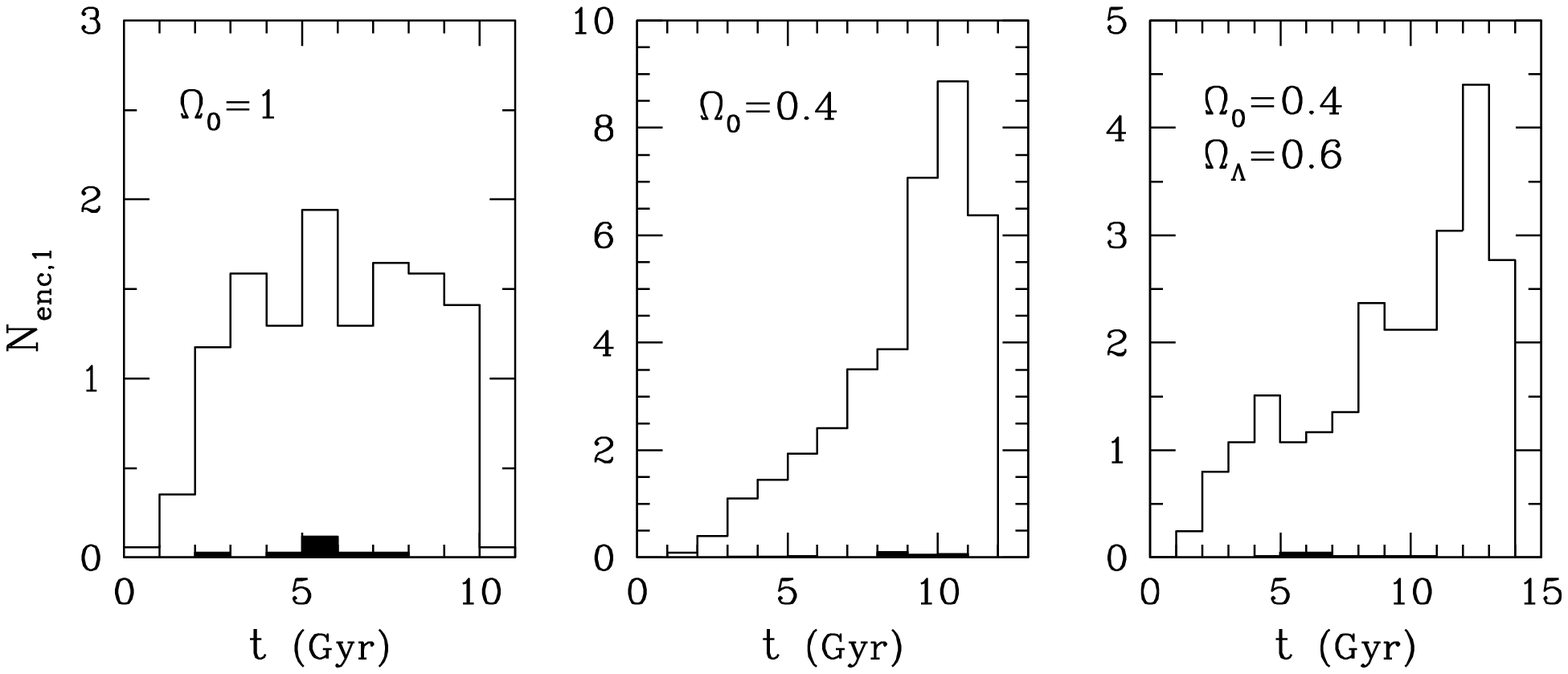}
\includegraphics[width=1.00\columnwidth,angle=0,clip]{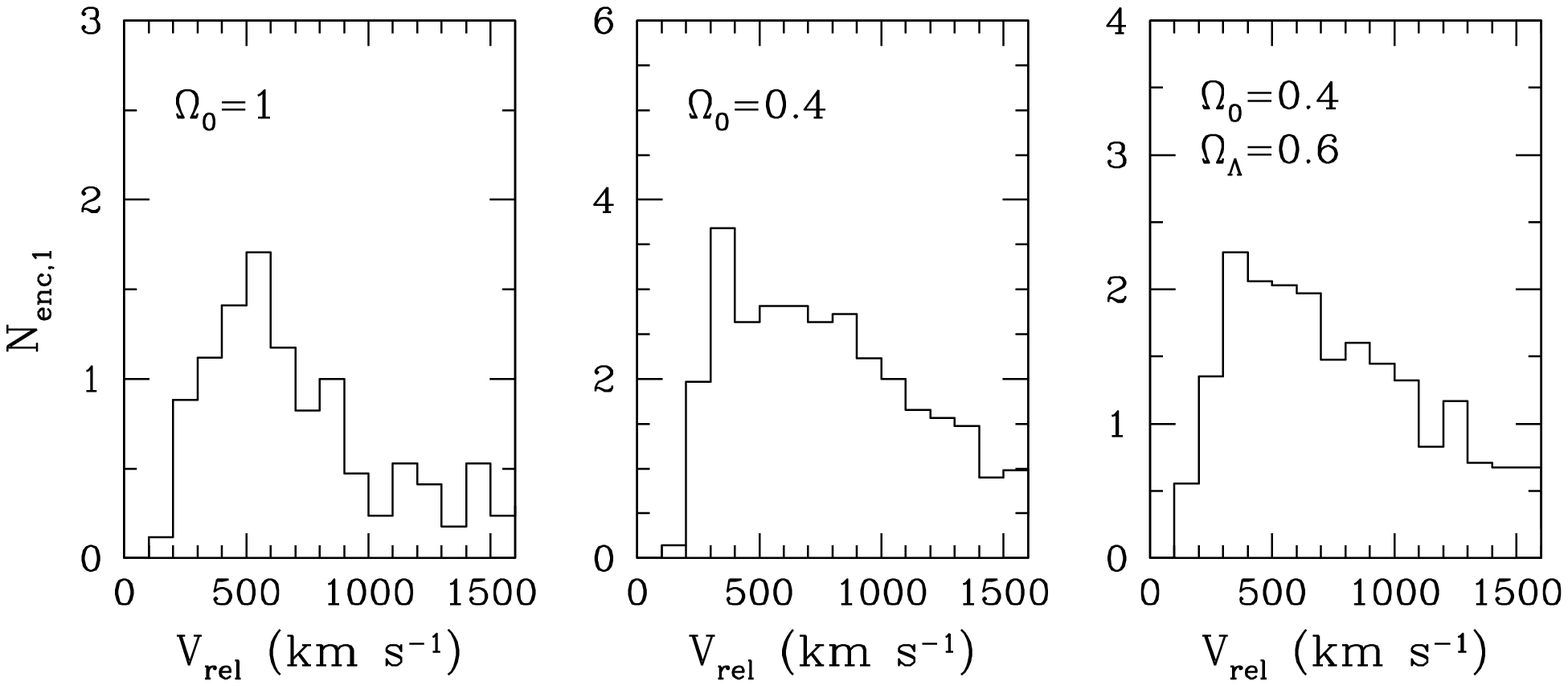}
\vskip 0pt \caption{
    Top: Number of close encounters per galaxy per Gyr in a
    simulated $\log (M/M_\odot) = 14.6$ cluster under different cosmologies.
    Bottom: The distribution of galaxy encounter velocities in the simulated
    clusters. (From Gnedin 2003.)
\label{gnedin}}
\end{figure*}

Even as the effects of high-speed collisions are being demonstrated, a
new attention is focusing on slow encounters and mergers in clusters.
The crucial element that is often overlooked in a classical
discussion of cluster interactions is the fact that structure forms
{\it hierarchically}. Galaxy clusters form not by accreting individual
galaxies randomly from the field environment, but rather through the
infall of less massive groups falling in along the filaments that make
up the ``cosmic web.'' Observationally, evidence for this accretion of
smaller galaxy groups is well established. Clusters show ample
evidence for substructure in X-rays, galaxy populations, and velocity
structure (see, \eg reviews by Buote 2002; Girardi \& Biviano 2002)
. These infalling groups have velocity dispersions that are much
smaller than that of the cluster as a whole, permitting the slow,
strong interactions normally associated with field galaxies. Imaging
of distant clusters show populations of strongly interacting and
possibly merging galaxies (\eg Dressler \etal 1997; van~Dokkum \etal
1999), which may contribute to the Butcher-Oemler effect (\eg Lavery
\& Henry 1988).  Even in nearby (presumably dynamically older)
clusters, several notable examples of strongly interacting systems
exist (\eg Schweizer 1998; Dressler, this volume), including the
classic Toomre-sequence pair ``The Mice'' (NGC 4476) located at a
projected distance of $\sim$5 Mpc from the center of the Coma
cluster. Interactions in the infalling group environment may in effect
represent a ``preprocessing'' step in the evolution of cluster
galaxies.

While the observational record contains many examples of group
accretion and slow encounters, numerical simulations are also
beginning to reveal this evolutionary path for cluster galaxies.
Modern $N$-body calculations can follow the evolution of individual
galaxy-mass dark matter halos in large-scale cosmological simulations,
allowing the interaction and merger history of cluster galaxies to be
probed. Ghigna \etal (1998) tracked galaxy halos in an $\Omega_m =1,
\log (M/M_\odot) = 14.7$ cluster and showed at late times ($z<0.5$) that, while
no mergers occurred within the inner virialized 1.5 Mpc of the
cluster, in the outskirts the merger rate was $\sim$5\%--10\%.
Similar models by Dubinski (1998) confirmed these
results, and showed more intense activity between $z=1$ and
$z=0.4$. Gnedin (2003) expanded on these works by studying the
interaction and merger rates in clusters under different cosmological
models (Fig. \ref{gnedin}). In $\Omega_m =1$ cosmologies, the
encounter rates in clusters stays relatively constant with time as the
cluster slowly accretes, while under open or $\Lambda$-dominated
cosmologies, the interaction rate increases significantly once the
cluster virializes, as the galaxies experience many high-speed
encounters in the cluster core.  The distribution of velocities in
these encounters shows a large tail to high encounter velocities, but
a significant fraction of encounters, largely those in the cluster periphery 
or those occurring at higher redshift, before the cluster has fully
collapsed, occur at low relative velocities ($\upsilon_{\rm rel} < 500$ \kms).
Clearly, not {\it all} interactions are of the high-speed variety!

In summary, clusters are an active dynamical environment, with a
multitude of processes available to drive evolution in the galaxy
population. Disentangling these different processes continues to prove
difficult, in large part because nearly all of them correlate with
cluster richness and clustercentric distance. Indeed, seeking to
isolate the ``dominant'' mechanism driving evolution may be
ill motivated, as these processes likely work in concert, such as the
connection between high-speed collisions and tidal field that
describes galaxy harassment. What is clear from both observational and
computational studies is that slow encounters and mergers of galaxies
can be important over the life of a cluster --- at early times when
the cluster is first collapsing, and at later times in the outskirts
as the cluster accretes groups from the field.

Finally, it is also particularly important to remember two
points. First, clusters come in a range of mass and richness: not
every cluster is as massive as the archetypal Coma cluster, with its
extraordinarily high velocity dispersion of $\sigma_\upsilon = 1000$ \kms. In
smaller clusters and groups, the lower velocity dispersion will slow
the encounter velocities and make them behave more like field
encounters. Second, as we push observations out to higher and higher
redshift, we begin to probe the regime of cluster formation, where
unvirialized dense environments can host strong encounters.

\section{Lessons from the Field ...}

\begin{figure*}[t]
\includegraphics[width=1.00\columnwidth,angle=0,clip]{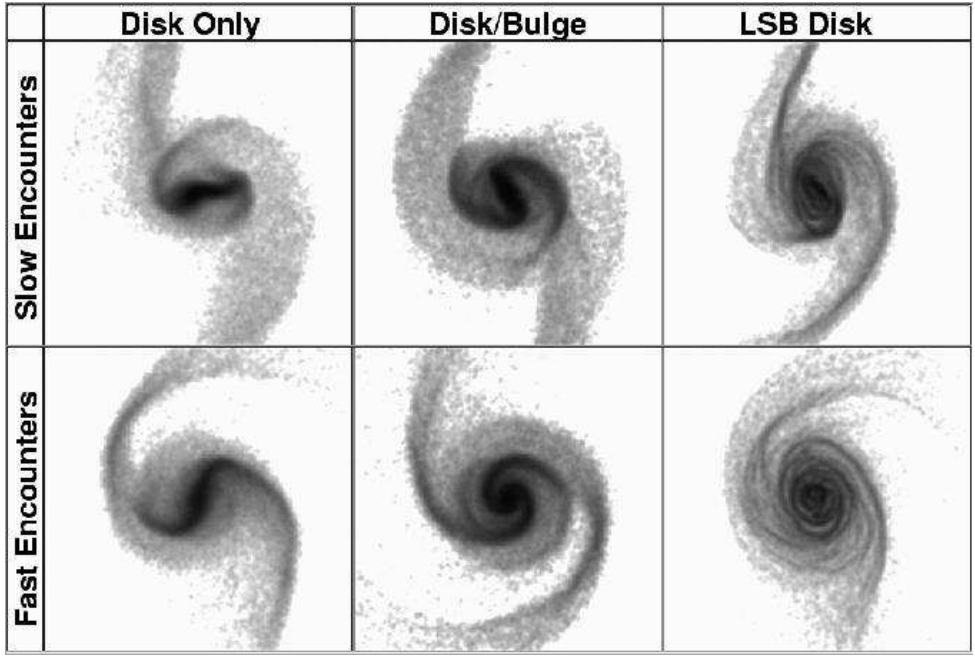}
\vskip 0pt \caption{
    The morphological response of galaxies to a close
    encounter.  The galaxy models are viewed one rotation period after
    the initial collision. Top panels show the response to a slow,
    parabolic encounter, while the bottom panels show the response to
    fast encounters. The left columns show the response of a pure disk
    system, the middle panels show a disk/bulge system, and the right panels
    show a low-density, dark matter dominated disk.
\label{collisions}}
\end{figure*}

To understand the effects of interactions and mergers on cluster
galaxies, we start with lessons learned from the study of collisions
in the field environment. In \S 1.3, we will then ask how the cluster
environment modifies these results. In this discussion, we focus
largely on two aspects of encounters: the triggering of starbursts and
nuclear activity, and the late-time evolution of tidal debris and
possible reformation of gaseous disks.

The role of galaxy interactions in driving activity and evolution of
field spirals has been well documented through a myriad of
observational and theoretical studies. Models of interactions have
demonstrated the basic dynamical response of galaxies to close
encounter (\eg Toomre \& Toomre 1972; Negroponte \& White 1982;
Barnes 1988, 1992; Noguchi 1988; Barnes \& Hernquist 1991, 1996; Mihos
\& Hernquist 1994, 1996). Close interactions can lead to a strong
internal dynamical response in the galaxies, driving the formation of
spiral arms and, depending on the structural properties of the disks,
strong bar modes. These non-axisymmetric structures lead to
compression and inflow of gas in the disks, elevating star formation
rates and fueling nuclear starburst/AGN activity. If the encounter is
sufficiently close, dynamical friction leads to an eventual merging of
the galaxies, at which time violent relaxation destroys the
dynamically cold disks and produces a kinematically hot merger remnant
with many of the properties found in the field elliptical galaxy
population (see, \eg Barnes \& Hernquist 1992).

Observational studies support much of this picture. Interacting
systems show preferentially elevated star formation rates, enhanced on
average by factors of a few over those of isolated spirals (Larson \&
Tinsley 1978; Condon \etal 1982; Keel \etal 1985; Kennicutt \etal
1987). Nuclear starbursts are common, with typical starburst mass
fractions that involve a few percent of the luminous mass (Kennicutt
\etal 1987). More dramatically, infrared-selected samples of 
galaxies reveal a population of interacting ``ultraluminous infrared
galaxies,'' where star formation rates are elevated by 1--2 orders of
magnitude and dust-enshrouded nuclear activity is common (Soifer
\etal 1984; Lawrence \etal 1989). These systems are preferentially
found in late-stage mergers (\eg Veilleux, Kim, \& Sanders 2002) and have been
suggested as the precursors of luminous quasars (Sanders \etal
1988). This diversity of properties for interacting systems argues
that the response of a galaxy to a close interaction is likely a
complicated function of encounter parameters, galaxy type, local
environment, and gas fraction.

Can we isolate the different determining factors to understand what
drives the strong response in interacting systems? Numerical modeling
of interactions has shown that gaseous inflow and central activity in
an interacting disk is driven largely by gravitational torques acting
on the gas --- not from the companion galaxy, but by the developing
non-axisymmetric structures (spiral arms and/or central bar) in the
host disk (Noguchi 1988; Barnes \& Hernquist 1991; Mihos \& Hernquist
1996). This result argues that the structural properties of galaxies
play a central role in determining the response to interactions.  In
particular, disks that are stable against the strong growth of disk
instabilities will experience a weaker response, exhibiting modestly
enhanced, disk-wide star formation (unless and until they ultimately
merge).  This stability can be provided by the presence of a centrally
concentrated bulge (Mihos \& Hernquist 1996) or a lowered disk surface
density (at fixed rotational speed; Mihos, McGaugh, \& de~Blok 1997).  In 
contrast, disk-dominated systems are more susceptible to global bar modes and
experience the strongest levels of inflow and nuclear activity (see
\S~1.3).

\begin{figure*}[t]
\includegraphics[width=1.00\columnwidth,angle=0,clip]{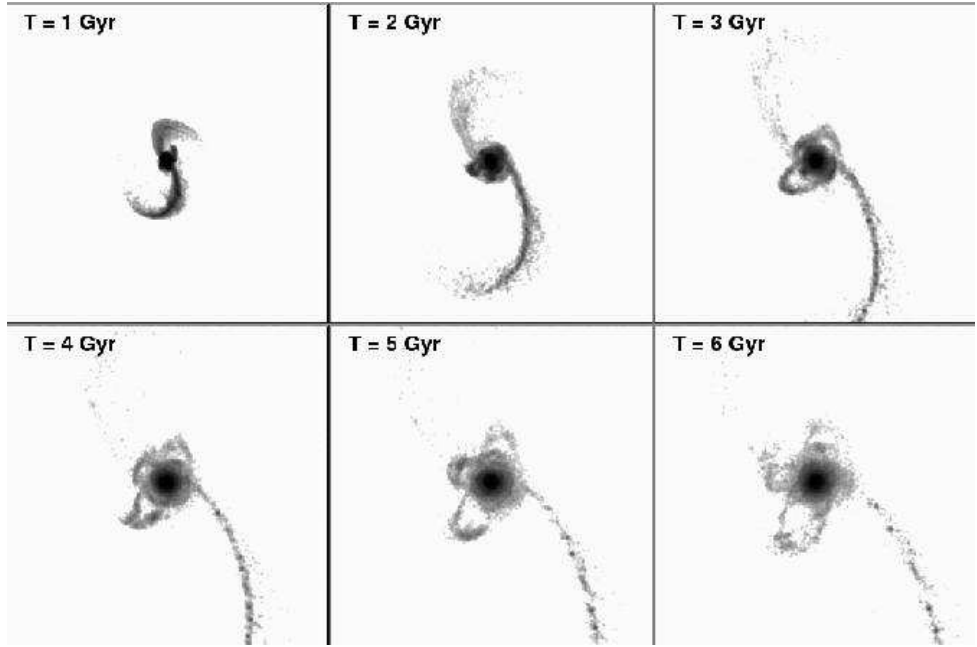}
\vskip 0pt \caption{
    Evolution of the tidal debris in an equal-mass merger of two disk galaxies
    occurring in isolation. Each frame is approximately 0.9 Mpc
    on a side.  Note the sharpness of the tidal debris, as
    well as the loops that form as material falls back into the
    remnant over long time scales.
\label{fieldmerger}}
\end{figure*}

Interacting and merging galaxies also show a wide variety of tidal
features, from long thin tidal tails to plumes, bridges, and other
amorphous tidal debris. The evolution of this material was first
elegantly described by the computer models of Toomre \& Toomre (1972) and
Wright (1972). Gravitational tides during a close encounter lead to
the stripping of loosely bound material from the galaxies (see
Fig.~\ref{fieldmerger}); rather than being completely liberated, the
lion's share of this material ($> 95$\%) remains bound, albeit
weakly, to its host galaxy (Hernquist \& Spergel 1992; Hibbard \&
Mihos 1995). Material is sorted in the tidal tails by a combination of
energy and angular momentum --- the outer portions of the tails
contain the least bound material with the highest angular
momentum. At any given time, material at the base of the tail has
achieved turn-around and is falling back toward the remnant. Further
out in the tail, material still expands away, resulting in a rapid
drop in the luminosity density of the tidal tails due to this
differential stretching. As a result, the detectability of these tidal
features is a strong function of age and limiting surface brightness;
after a few billion years of dynamical evolution, they will be
extremely difficult to detect (Mihos 1995).

In mergers, the gas and stars ejected in the tidal tails fall back
onto the remnant in a long-lived ``rain'' that spans many billions of
years (Hernquist \& Spergel 1992; Hibbard \& Mihos 1995).  In the
merger simulation shown in Figure~\ref{fieldmerger}, this fallback
manifests itself as loops of tidal debris that form as stars fall
back through the gravitational potential of the remnant.  Tidal gas
will follow a different evolution, as it shocks and dissipates energy
as it falls back.  The most tightly bound gaseous material returns to
the remnant over short time scales and can resettle into a warped disk
(Mihos \& Hernquist 1996; Naab \& Burkert 2001; Barnes 2002).  Such warped
H~I disks have been observed in NGC 4753 (Steiman-Cameron, Kormendy, \& Durisen
1992) and in the nearby merger remnant Centaurus A (Nicholson, Bland-Hawthorn, 
\& Taylor 1992).  Over longer time scales, the loosely bound, high-angular 
momentum gas falls back to ever-increasing radii, forming a more extended but
less-organized distribution of gas outside several effective radii in
the remnant. Many elliptical galaxies show extended neutral hydrogen
gas, sometimes in the form of broken rings at large radius, perhaps
arising from long-ago merger events (\eg van~Gorkom \& Schiminovich 1997).

The ultimate fate of this infalling material is uncertain, but may
have important ramifications for interaction-driven galaxy evolution
models.  If efficient star formation occurs in this gas, such as
that observed in the inner disk of the merger remnant NGC 7252
(Hibbard \etal 1994), this may present a mechanism for building
disks in elliptical galaxies. If the amount of gas resettling into the
disk is significant, in principle the remnant could evolve to become a
spheroidal system with a high bulge-to-disk ratio, perhaps forming an
S0 or Sa galaxy (\eg Schweizer 1998).

\section{... Applied to Clusters}

In clusters, a number of environmental effects may modify the
dynamical response and evolution of interacting galaxies described
above. First, the relative velocities of interacting systems tend to
be higher, although, as argued earlier, many low-velocity encounters
still occur within smaller groups falling in from the cluster
periphery. Second, the global tidal field of the cluster must also
play a role in the evolution of interacting systems, stripping away
the loosely bound tidal material and potentially adding energy to
bound groups. The hot intracluster medium (ICM) can act to further strip out
low-density ISM in galaxies, particularly the diffuse
tidal gas ejected during collisions.

The most obvious difference between interactions in the field and in
the cluster environment is the collision speed of the encounter. While
slow interactions are able to drive a strong dynamical response in
disk galaxies, faster encounters result in a perturbation that is
much shorter lived and less resonant with the internal dynamics of the
disk. Figure~\ref{collisions} shows the different response of galaxies
to slow and fast collisions. In each case, the galaxies experience an
equal-mass encounter inclined 45$^\circ$ to the orbital plane and with
a closest approach of six disk scale lengths. Three structural models
are used for the galaxies. The first is a pure disk system where the
disk dominates the rotation curve in the inner two scale lengths; the
second is identical to the first, but with a central bulge with
bulge-to-disk ratio of 1:3. In both these models, the disk-to-halo mass
ratio is 1:5.8. The third system is identical to the first, but with
the disk surface density lowered by a factor of 8; this system
represents a dark matter dominated, low-surface brightness (LSB) disk
galaxy. In the slow collision, the galaxies fall on a parabolic (zero-energy) 
orbit with a velocity at closest approach that is
approximately twice the circular velocity. The fast collision takes
place with a hyperbolic orbit with an encounter velocity of twice that
of the parabolic encounters.

Notable differences can be seen in the dynamical response of the
galaxy models, particularly in the growth of global bar modes. During
slow encounters, both the disk and disk/bulge galaxy models develop
dramatic bars and spiral arms, which can drive strong inflow and
central activity. The lowered surface density of the LSB model results
in a weaker self-gravitating response (Mihos \etal 1997); a very
small, weak bar is present, but the overall response is one of a
persistent oval distortion, which would be much less able to drive
gaseous inflow.  In contrast, the response of the fast encounters
depends more strongly on the structural properties of the galaxy. The
pure-disk model develops a relatively strong bar mode, while the
disk/bulge system sports a two-arm spiral pattern with no central
bar. These results are similar to those shown in Moore \etal (1999),
who modeled high-speed encounters of disk galaxies of varying
structural properties. The LSB model, on the other hand, lacks the
disk self-gravity to amplify the perturbation into any strong internal
response. However, the vulnerability of LSBs lies not in their
internal response to a {\it single} encounter, but rather in their
response to {\it repeated} high-speed collisions in the cluster
environment (Moore \etal 1998).

Because the star-forming response of a galaxy is intimately linked to
its dynamical response, we can use these results to guide our
expectations of starburst triggering mechanisms in cluster galaxies.
Because of their stability toward high-speed encounters, luminous,
early-type spirals should experience modestly enhanced, disk-wide
starbursts. Low-luminosity, late-type disks will be more susceptible to
stronger inflows, central starbursts, and AGN fueling; even the LSBs
will succumb to the effects of repeated encounters and the
cluster tides (Moore \etal 1996), which drive a much stronger
response. As a result, these high-speed, ``harassment-like''
encounters are effective at driving evolution in the low-luminosity
cluster populations (Moore \etal 1996; Lake \etal 1998), but if
harassment is the whole story in driving cluster galaxy evolution, it
is hard to explain strong starburst activity in {\it luminous} cluster
spirals at moderate redshift. On the other hand, the slower collisions
expected in infalling substructure are able to drive a stronger
response regardless of galaxy type.

\begin{figure*}[t]
\includegraphics[width=1.00\columnwidth,angle=0,clip]{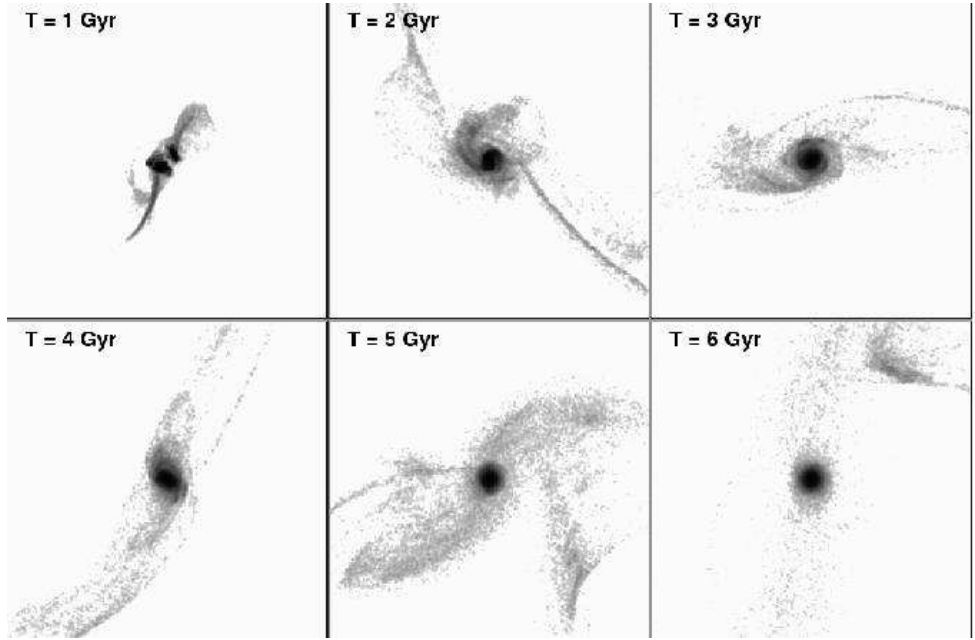}
\vskip 0pt \caption{
    Evolution of an equal-mass merger, identical to that in
    Fig.~\ref{fieldmerger}, but occurring as the system orbits through a
    Coma-like cluster potential (see text).  Note the rapid stripping of
    the tidal tails early in the simulation; the tidal debris seen
    here is more extended and diffuse than in the field merger, and
    late infall is shut off due to tidal stripping by the cluster
    potential.
\label{clustermerger}}
\end{figure*}

Aside from driving stronger starbursts in interacting cluster
galaxies, slow collisions also heat and strip galaxies more
efficiently than do high-speed encounters. They also raise the
possibility for mergers among cluster galaxies, and the potential
for merger-driven evolutionary scenarios. However, unlike slow
collisions in the field, cluster galaxies must also contend with the
effects of the overall tidal field of the cluster. How will this
affect the evolution of close interactions, in particular the
longevity and detectability of tidal debris, and the ability for
galaxies to reaccrete tidal material?  To address this question,
Figure~\ref{clustermerger} shows the evolution of an equal-mass merger
(identical to the merger shown in Fig.~\ref{fieldmerger})
occurring in a cluster tidal field. The cluster potential is given by
an Coma-like Navarro, Frenk, \& White (1996) profile with total mass 
$M_{200} = 10^{15}\,M_\odot$,
$r_{200} = 2$ Mpc, and $r_s = 300$ kpc. The binary pair travels on an
orbit with $r_{\rm peri} = 0.5$ Mpc and $r_{\rm apo} = 2$ Mpc, and passes
through periclustercon twice, at T $\approx$ 1 and 4 Gyr.

The tidal field has a number of effects on the evolution of this
system. First, the merger time scale has been lengthened --- the
cluster tidal field imparts energy to the galaxies' orbits, extending
the time it takes for the system to merge (by $\sim$50\% for this
calculation). In this case, the encounter is close enough that the
galaxies do still ultimately merge, but it is not hard to envision
encounters where the tidal energy input is sufficient to unbind the
galaxy pair. This raises the interesting possibility that infalling
pairs may experience the close, slow collisions that drive strong
activity, yet survive the encounter whole without merging.

The most dramatic difference between field mergers and those in
a cluster is in the evolution of the tidal debris. In the field, most of the
material ejected into the tidal tails remains loosely bound to the
host galaxy, forming a well-defined tracer of the tidal encounter. In
the cluster encounter, the tidal field quickly strips this loosely
bound material from the galaxies, dispersing it throughout the cluster
and adding to the diffuse intracluster starlight found in galaxy
clusters. This rapid stripping of the tidal debris is clear in
Figure~\ref{clustermerger}: shortly after the first passage the debris
becomes very extended and diffuse and is removed entirely from the
system after the subsequent passage through the cluster interior.  In
this encounter, 20\% of the stellar mass is stripped to large
distances ($>$ 50 kpc), fully twice the amount in a similar field
merger.

This rapid stripping has a number of important ramifications. First,
these tidal tracers are very short-lived; identifying a galaxy as a
victim of a close interaction or merger will be very difficult indeed
shortly after it enters the cluster potential. The tidal features we
do see in cluster galaxies are likely signatures of a very recent
interaction, such that interaction rates derived from the presence of
tidal debris may underestimate the true interaction rates in
clusters. Second, the rebuilding/resettling of gaseous disks in
interaction/merger remnants will be severely inhibited, as both
cluster tides and ram pressure stripping act to strip off all but the
most bound material in the tidal debris, leaving little material able
to return. Finally, this stripping will contribute both to the
intracluster light and to the ICM, as gas and stars in
the tidal debris are mixed in to the diffuse cluster environment. We
discuss these processes in the following sections.

\section{Galaxy Evolution: Mergers, Elliptical, and S0 Galaxies}

In the field environment, interactions and mergers of galaxies are
thought to be a prime candidate for driving evolution in galaxy
populations. Major mergers of spiral galaxies may lead to the
transformation of spirals into ellipticals (\eg Toomre 1978; Schweizer
1982): the violent relaxation associated with a merger effectively
destroys the galactic disks and creates a kinematically hot,
$r^{1/4}$-law spheroid, while the concurrent intense burst of star
formation may process the cold ISM of a spiral galaxy into the hot
X-ray halo of an elliptical. These processes are not totally
efficient, however, and leave signatures behind that identify the
violence of the merging process: diffuse loops and shells of
starlight, extended H~I gas, significant rotation in the outskirts of
the remnant, and dynamically distinct cores (see, e.g., the
review by Schweizer 1998). While it is an open question as to what
fraction of ellipticals formed this way, it is clear that at least
some nearby field ellipticals --- Centaurus~A being a notable example
(Schiminovich \etal 1994) --- have had such violent histories.

Mergers have also been proposed as a mechanism to drive the formation
of S0 galaxies. In this case, the scenarios are varied. For
S0s with very large bulge-to-disk ratios, reaccretion of gas after a
major merger (either from returning tidal material or from the
surrounding environment) may rebuild a disk inside a newly formed
spheroid. In unequal-mass mergers of disk galaxies, disk destruction
is not complete, and the resulting remnant retains a significant
amount of rotation (Bendo \& Barnes 2000; Cretton \etal 2001) and may
be identified with a disky S0, particularly if a significant amount of
cold gas is retained by the system to reform a thin disk (Bekki 1998). 
Finally, minor mergers between spirals and their satellite companions can 
significantly heat, but not destroy, galactic disks (Toth \& Ostriker 1992; 
Quinn, Hernquist, \& Fullagar 1993; Walker, Mihos, \& Hernquist 1996), while
simultaneously helping to ``sweep the disk clean'' of cold gas via a
gravitationally induced bar driving gas to the nucleus (Hernquist \&
Mihos 1995). The resulting disks have many similarities to disky S0s
(Mihos \etal 1995): thickened disks, little or no spiral structure,
cold gas, or ongoing star formation. These different scenarios vary
mainly in the proposed strength of the interaction --- from major
to minor mergers --- and it has been proposed that this parameter may,
in fact, determine the ultimate morphological classification of
galaxies all the way from early-type ellipticals to late-type spirals
(\eg Schweizer 1998; Steinmetz \& Navarro 2002).

Applying these arguments to cluster populations, it seems that
building cluster ellipticals through a wholesale merging of spirals
within the established cluster environment is a difficult
proposition. Clusters ellipticals are an old, homogeneous population
showing little evolution since at least a redshift of $z\approx 1$ (\eg
Dressler \etal 1997; Ellis \etal 1997). Within the cores of massive
clusters, merging has largely shut off due to the high velocity
dispersion of the virialized cluster (Ghigna \etal 1998). The
accretion of merger-spawned ellipticals from infalling groups may
still occur, and these will be hard to identify morphologically as
merger remnants --- the combination of cluster tides and hot ICM will
strip off any tell-tale tidal debris and sweep clean any diffuse cold
gas in the tidal tails (recall Fig.~\ref{clustermerger}) or low-density 
reaccreting disk. However, the small scatter in the
color-magnitude relation and weak evolution of the fundamental plane
of cluster ellipticals (see, \eg the review by van~Dokkum 2002) argues
that such lately formed ellipticals likely do not contribute to the bulk
of the cluster elliptical population. This does not mean that mergers
have not played a role in the formation of cluster ellipticals. In any
hierarchical model for structure formation, galaxies form via the
accretion of smaller objects.  Luminous cluster ellipticals may well
have formed from mergers of galaxies at high redshift, in the
previrialized environment of the protocluster.  However, at these
redshifts ($z \gg 1$), the progenitor galaxies are likely to have looked
very different from the present-day spiral population.

Unlike the rather passive evolution observed in cluster ellipticals,
much stronger evolution is observed in the population of cluster S0s.
The fraction of S0s in rich cluster has increased significantly
since a redshift of $z\approx 1$, with a corresponding decrease in the
spiral fraction (Dressler \etal 1997).  Can the same collisional
processes that have been hypothesized to drive S0 formation in the
field account for the dramatic evolution in cluster S0 populations?
S0 formation scenarios that rely on reaccretion of material after a
major merger (disk rebuilding schemes) seem difficult to envision
wholly within the cluster environment. While mergers are possible in
infalling groups, the combination of tidal and ram pressure stripping
will shut down reaccretion and ablate any low-density gaseous disks
that have survived the merger process. For example, it is unlikely the
H~I disk in Centaurus A (Nicholson \etal 1992), likely a product of
merger accretion, would survive passage through the hot ICM of a
dense cluster. Satellite merger mechanisms trade one dynamical problem
for another --- because the mergers involve bound satellite
populations there is no concern about the efficacy of high-speed
mergers, but, instead, the issue is whether or not satellite populations
can stay bound to their host galaxy as it moves through the cluster
potential. And, of course, this mechanism relies on the very {\it local}
environment of galaxies, which does not explain why S0 formation would be
enhanced in clusters.

None of the proposed merger-driven S0 formation mechanisms appear to
work well deep inside the cluster potential. On the other hand, these
processes should operate efficiently in the group environment, where
the encounter velocities are smaller and cluster tides and the  hot ICM do
not play havoc with tidal reaccretion. The group environment may
create S0s and feed them into the accreting cluster, but if there is
wholesale transformation of cluster spirals into S0s in the cluster
environment, it needs to occur via other mechanisms.

Other cluster-specific methods for making S0 galaxies have been proposed, 
including collisional heating and ram pressure stripping of the dense ISM 
(Moore \etal 1999; Quilis, Moore, \& Bower 2000) and strangulation, the 
stripping of hot halo gas from spirals (Larson \etal 1980;
Bekki, Couch, \& Shioya 2002). While these models, by design, explain the
preferential link between clusters and S0 galaxies, they are not
without problems themselves. While the effects of ram pressure
stripping on the extended neutral hydrogen gas in cluster galaxies is
clear (\eg van~Gorkom, this volume), its efficacy on the denser
molecular gas is unclear. For example, the H~I deficient spirals in the
Virgo cluster still contain significant quantities of molecular gas
(\eg Kenney \& Young 1989), while studies of the molecular content of
cluster spirals show no deficit of CO emission (Casoli \etal 1998). If
the molecular ISM survives, it is unclear why star formation should
not continue in these disks.  Strangulation models suffer less from concerns 
of the efficacy of ram pressure stripping, since it is much easier to strip 
low-density halo gas than a dense molecular ISM, although it must be noted 
that there currently is little observational evidence for hot halos in
(non-starbursting) spiral galaxies. In addition, neither of these
methods leads to the production of a luminous spheroid --- the S0s
that might be produced in these ways would have low bulge-to-disk
ratios.

Ultimately, S0s are a heterogeneous class, from bulge-dominated S0s
to the disky S0s seen in galaxy clusters, and it should not be
surprising that a single mechanism cannot fully account for the range
of S0 types (\eg Hinz, Rix, \& Bernstein 2001). Whether there is a systematic
difference between cluster and field S0s is unclear, an issue fraught with
selection and classification uncertainties. What is clear is that,
even in clusters, S0s often show evidence for accretion events,
similar to that observed in the field S0 population (see, \eg, the
discussion in Schweizer 1998). It is likely that many of these S0s
were ``processed'' via mergers in the group environment before being
incorporated into clusters.

\section{Tidal Stripping and Intracluster Light}

As galaxies orbit in the potential well of a galaxy cluster, stars are
tidally stripped from their outer regions, mixing over time to form a
diffuse ``intracluster light'' (ICL). First proposed by Zwicky (1951),
the ICL has proved very difficult to study --- at its {\it brightest},
it is only $\sim$1\% of the brightness of the night sky. Previous
attempts to study the ICL have resulted in some heroic detections (Oemler
1973; Thuan \& Kormendy 1977; Bernstein \etal 1995; Gregg \& West 1998; 
Gonzalez \etal 2000), verified by observations of intracluster stars and 
planetary nebulae in Virgo (Feldmeier, Ciardullo, \& Jacoby 1998; Ferguson, 
Tanvir, \& von~Hippel 1998; Arnaboldi \etal 2002). While the ICL is typically 
thought of as arising from the stripping of starlight due to the cluster
potential, in fact the role of interactions between cluster galaxies
in feeding the ICL is quite strong.  Galaxy
interactions significantly enhance the rate at which material is
stripped; as illustrated in Figure~\ref{clustermerger}, the strong,
local tidal field of a close encounter can strip material from deep
within a galaxy's potential well, after which the cluster tidal field
can liberate the material completely. Interactions, particularly those
in infalling groups, act to "prime the pump" for the creation of the
ICL.

The properties of the ICL in clusters, particularly the fractional
luminosity, radial light profile, and presence of substructure, may
hold important clues about the accretion history and dynamical
evolution of galaxy clusters. Material stripped from galaxies falling
in the cluster potential is left on orbits that trace the orbital
path of the accreted galaxy, creating long, low-surface brightness
tidal arcs (\eg Moore \etal 1996), which have been observed in a few
nearby clusters (Trentham \& Mobasher 1998; Calc\'aneo-Rold\'an \etal\
2000). However, these arcs will only survive as discrete structures if
the potential is quiet; substructure will dynamically heat these arcs,
and the accretion of significant mass (\ie a cluster merger event) may
well destroy these structures. If much of the ICL is formed early in a
cluster's dynamical history, before the cluster has been fully
assembled, the bulk of the ICL will be morphologically smooth and
well mixed by the present day, with a few faint tidal arcs showing the
effects of late accretion. In contrast, if the ICL formed largely
after cluster virialization, from the stripping of ``quietly
infalling'' galaxies, the ICL should consist of an ensemble of
kinematically distinct tidal debris arcs. Clusters that are
dynamically younger should also possess an ICL with significant
kinematic and morphological substructure.

Early theoretical studies of the formation of the ICL
suggested that it might account for anywhere from 10\% to 70\%
of the total cluster luminosity (Richstone 1976; Merritt 1983, 1984; Miller
1983; Richstone \& Malumuth 1983; Malumuth \& Richstone 1984). These studies 
were based largely on analytic estimates of tidal
stripping, or on simulations of individual galaxies orbiting in a
smooth cluster potential well.  Such estimates miss the effects of
interactions with individual galaxies (\eg Moore \etal 1996),
intermediate-scale substructure (Gnedin 2003), and priming due to
interactions in the infalling group environment. As a result, these
models underpredict the total amount of ICL as well as
the heating of tidal streams in the ICL. Now, however, cosmological
simulations can be used to study cluster collapse and tidal stripping
at much higher resolution and with a cosmologically motivated cluster
accretion history (\eg Moore \etal 1998; Dubinski, Murali, \& Ouyed 2001).

\begin{figure*}[t]
\includegraphics[width=1.00\columnwidth,angle=0,clip]{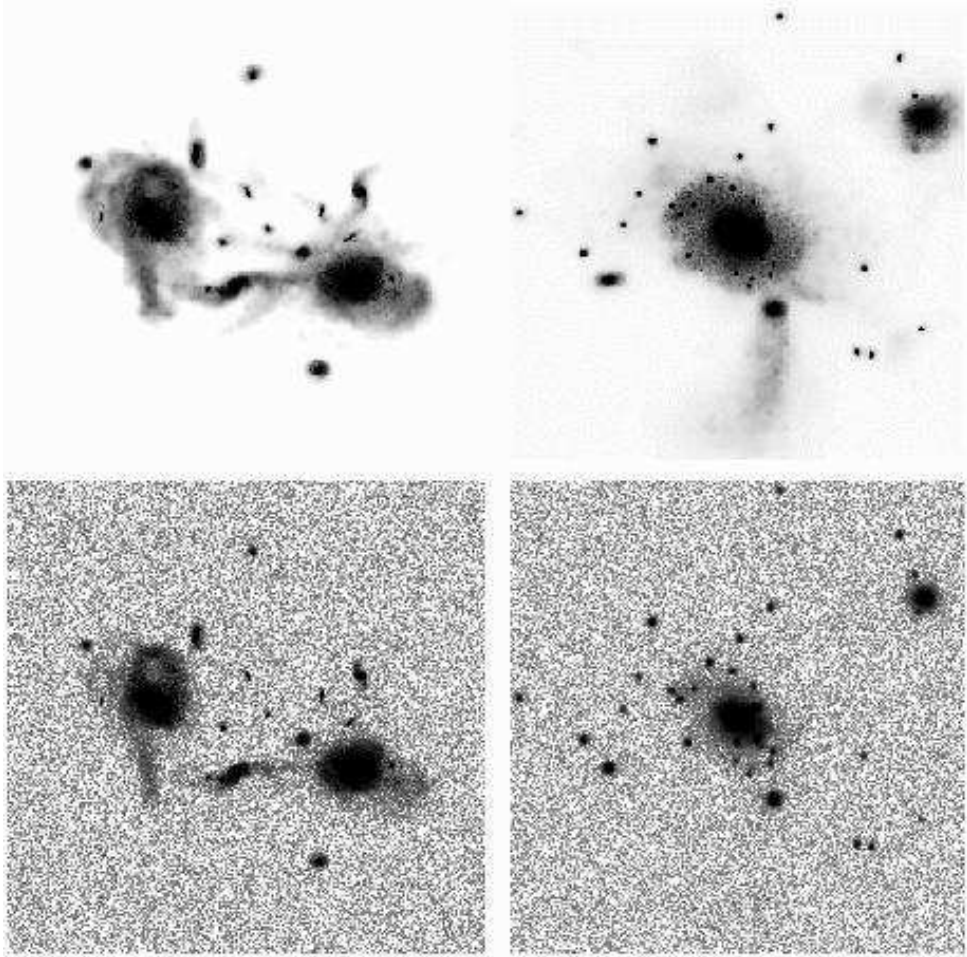}
\vskip 0pt \caption{
    Visualizations of the cluster simulations of Dubinski (1998). Top panels 
    show the distribution of luminous starlight, with the faintest contours 
    corresponding to a surface brightness of $\mu_V \approx 30$ mag 
    arcsec$^{-2}$. Bottom panels show the effect of adding noise
    characteristic of current observational limits. Left panels show
    the cluster early in collapse (at $z=2$), while right panels show
    the virialized cluster at $z=0$.
\label{iclmodel}}
\end{figure*}

One example of the modeling of ICL is shown in Figure~\ref{iclmodel}.  These 
images are derived from the $N$-body simulations of Dubinski (1998), who 
simulated the collapse of a $\log (M/M_\odot) = 14.0$ cluster in a standard 
cold dark matter Universe.  Starting from a
cosmological dark matter simulation, the 100 most massive halos are
identified at a redshift of $z=2.2$ and replaced with composite
disk/bulge/halo galaxies, whereafter the simulation is continued to
$z=0$ (see Dubinski 1998 for more details). To quantify the diffuse
light in these cluster models, we assign luminosity to the stellar
particles based on a mass-to-light ratio of 1. The top panels show the
cluster at two different times. On the left, the cluster is shown
early in the collapse, at $z=2$, where it consists of two
main groups coming together. The right panels show the cluster at 
$z=0$, when the cluster has virialized and formed a
massive cD galaxy at the center. In each case, the lowest visible
contour is at a surface brightness of $\mu_V \approx 30$ mag arcsec$^{-2}$. 
The bottom panels show the effects of adding
observational noise typical of our ICL imaging data (discussed below)
and illustrate the difficulties in detecting this diffuse light.

\begin{figure*}[t]
\includegraphics[width=1.00\columnwidth,angle=0,clip]{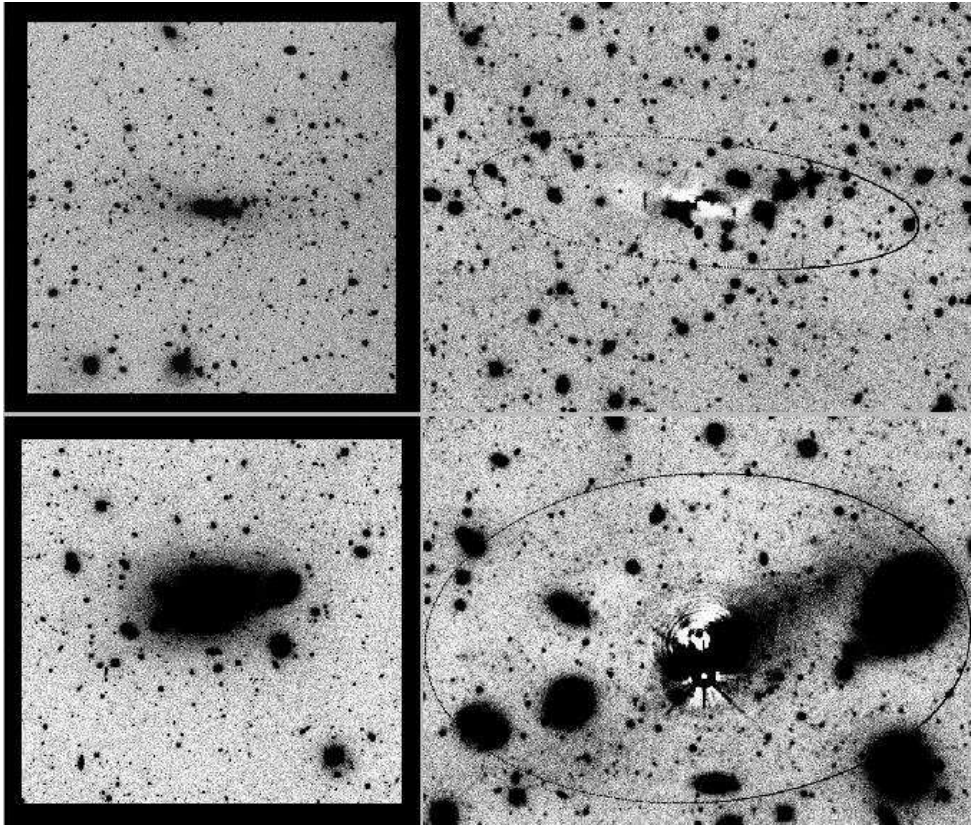}
\vskip 0pt \caption{
    Deep imaging of cD clusters from Feldmeier \etal (2002). Top panels show
    Abell 1413; bottom panels show MKW 7. The left panels show the full
    $10^\prime \times 10^\prime$ view of the cluster, while
    the right panels show a close up of the clusters once a smooth
    elliptical fit to the cD cluster envelope has been removed. The
    oval shows the radius inside which the model has been subtracted.
\label{cDclusters}}
\end{figure*}

\begin{figure*}[t]
\includegraphics[width=1.00\columnwidth,angle=0,clip]{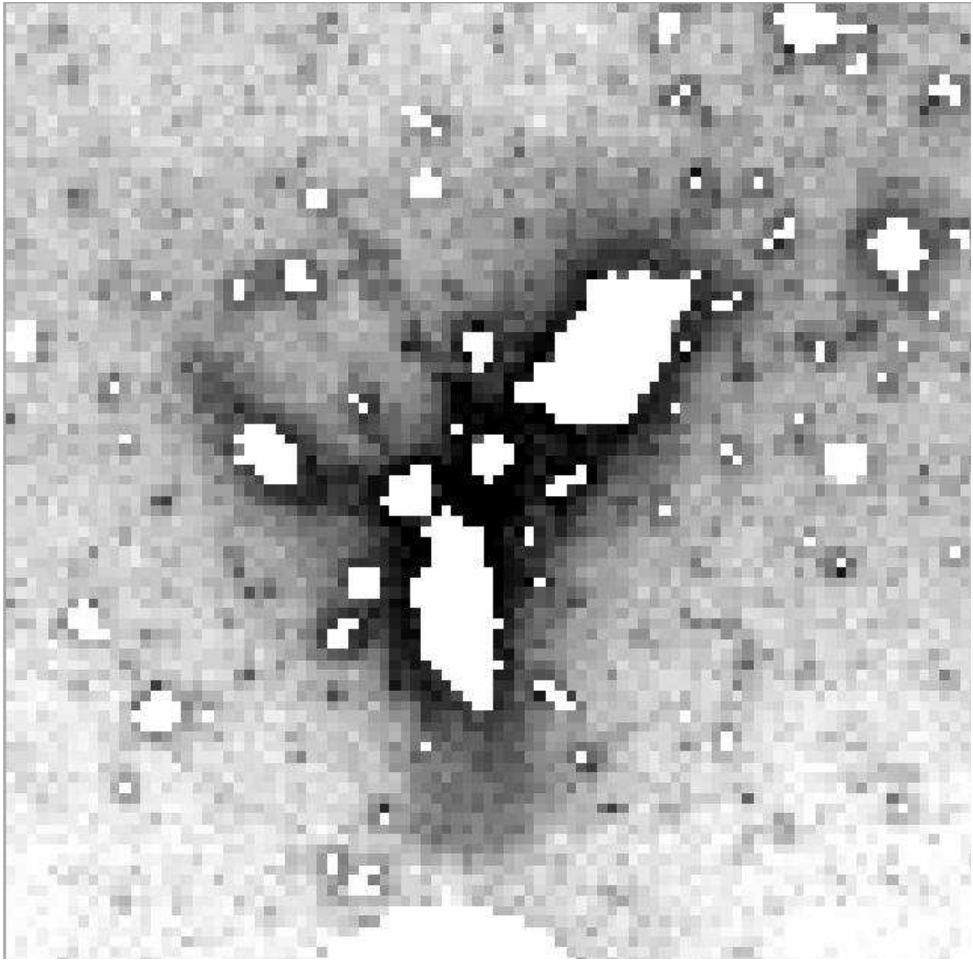}
\vskip 0pt \caption{
    The ICL in Abell 1914 (from Feldmeier \etal 2003).
\label{1914}}
\end{figure*}

In the early stages of cluster collapse, material is being stripped
out of galaxies and into the growing ICL component. This material has
a significant degree of spatial structure in the form of thin streams
and more diffuse plumes, much of it at observationally detectable
surface brightnesses. At later times this material has become
well mixed in the virialized cluster, forming a much smoother
distribution of ICL and substructure that is visible only at much
fainter surface brightnesses, well below current levels of
detectability. Along these lines, the degree of ICL substructure may
act as a tracer of the dynamical age of galaxy clusters.

Indeed, galaxy clusters do show a range of ICL properties. We
(Feldmeier \etal 2002, 2003) have recently begun a deep imaging survey
of galaxy clusters, aimed at linking their morphological properties to
the structure of their ICL. As the detection of ICL is crucially
dependent on reducing systematic effects in the flat fielding, we have
taken significant steps to alleviate these issues, including imaging
in the Washington M filter to reduce contamination from variable night
sky lines, flat fielding from a composite of many night sky flats
taken at similar telescope orientations, and aggressive masking of
bright stars and background sources (see Feldmeier \etal 2002 for
complete details). With this data, we achieve a signal-to-noise ratio of 5
at $\mu_V = 26.5$ mag arcsec$^{-2}$ and a signal-to-noise ratio of 1 at 
$\mu_V = 28.3$ mag arcsec$^{-2}$. We
have targeted two types of galaxy clusters thus far: cD-dominated
Bautz-Morgan class I clusters (Feldmeier \etal 2002) and irregular
Bautz-Morgan class III clusters (Feldmeier \etal 2003).

Figure~\ref{cDclusters} shows results for the cD clusters Abell 1413
and MKW 7.  Similar to the clusters studied by Gonzalez \etal (2000; see 
also Gonzalez, Zabludoff, \& Zaritsky 2003),
the cD galaxies are well fit by a $r^{1/4}$ law
over a large range in radius, with only a slight luminosity excess in
the outskirts of each cluster.  In each case, we search for ICL
substructure by using the STSDAS {\tt ellipse} package to subtract a
smooth fit to the cD galaxy extended envelope. In the case of Abell
1413, we see little evidence for any substructure in the ICL; the
small-scale arcs we observe are likely to be due to gravitational
lensing. MKW 7 shows a broad plume extending from the cD galaxy to a
nearby bright elliptical, but little else in the way of substructure.

In contrast, we see evidence for more widespread ICL substructure in
our Bautz-Morgan type III clusters. Figure~\ref{1914} shows our image
of Abell 1914, binned to a resolution of $3''$ after all stars and
galaxies have been masked. Here we see a variety of features: a
fan-like plume projecting from the southern clump of galaxies, another
diffuse plume extending from the galaxy group to the east of the
cluster, and a narrow stream extending to the northeast from the
cluster center. The amount of substructure seen here is consistent
with an unrelaxed cluster experiencing a merger, similar to the
features seen in the unrelaxed phase of the model cluster shown in
Figure~\ref{iclmodel}. We see similar plumes in other type III
clusters, suggesting that the ICL in these types of clusters does
reflect a cluster that is dynamically less evolved than the cD-dominated 
clusters of Feldmeier \etal (2002).

While these studies point toward significant substructure in the ICL
of galaxy clusters, imaging surveys continue to be hampered by
systematic effects.  With so much of the ICL substructure present
only at surface brightnesses fainter than $\mu_V > 28$ mag arcsec$^{-2}$, 
issues of flat fielding, scattered light, and sky variability become severe. An
interesting alternative is to use the significant numbers of
intracluster planetary nebulae now being found in emission-line
surveys of nearby galaxy clusters (Feldmeier \etal 1998; Arnaboldi
\etal 2002). These studies have very different detection biases than 
deep surface photometry and have the potential to probe the ICL down
to much lower surface densities. Planetary nebulae offer an added bonus: as
emission-line objects, follow-up spectroscopy can determine the
kinematics of the ICL, giving yet another view of the degree to which
the ICL is dynamically relaxed (Dubinski \etal 2001; Willman 2003).
An interesting analogy can be made between the search for
kinematic substructure due to tidal stripping in galaxy clusters and
the search for kinematic substructure due to tidally destroyed
satellites in the Milky Way's halo (\eg Morrison \etal 2002).  In both
cases, kinematic substructure can be used to trace the dynamical
accretion history of the system.  With the advent of multi-object
spectrographs on 8-m class telescopes, new and exciting opportunities
now exist for studying this substructure in the diffuse starlight of
galaxy clusters.

\vspace{0.3cm}
{\bf Acknowledgements}.
I would like to thank all my collaborators for their contributions to this
work, and John Feldmeier in particular for providing
Figures~\ref{iclmodel} -- \ref{1914}.  John Dubinski graciously
provided the cluster simulations from which Figure~\ref{iclmodel} was
made. I also thank the conference organizers for putting together such
a wonderful scientific program. 
This research has been supported in part by an NSF Career Award and a
Research Corporation Cottrell Scholarship.

\end{document}